\documentclass[amsmath,amssymb,aps,prx,reprint,superscriptaddress]{revtex4-2}

\usepackage{amsmath}
\usepackage{xcolor}
\usepackage{amsfonts,amssymb}
\usepackage{graphicx}
\usepackage{hyperref}
\usepackage{academicons}
\usepackage{multirow}
\usepackage{orcidlink}
\usepackage{booktabs}
\usepackage{wasysym}
\usepackage{needspace}

\begin{document}

\title{Improving robustness and training efficiency of machine-learned potentials by incorporating short-range empirical potentials}
\author{Zihan Yan\,\orcidlink{0000-0002-8911-6549}}
\affiliation{School of Materials Science and Engineering, Zhejiang University, Hangzhou, Zhejiang 310027, China}
\affiliation{Research Center for Industries of the Future and School of Engineering, Westlake University, 600 Dunyu Road, Hangzhou, Zhejiang 310030, China}

\author{Zheyong Fan\,\orcidlink{0000-0002-2253-8210}}
\affiliation{College of Physical Science and Technology, Bohai University, Jinzhou 121013, China}

\author{Yizhou Zhu\,\orcidlink{0000-0002-5819-7657}}
\email[]{zhuyizhou@westlake.edu.cn}
\affiliation{Research Center for Industries of the Future and School of Engineering, Westlake University, 600 Dunyu Road, Hangzhou, Zhejiang 310030, China}

\date{\today}

\begin{abstract}
    
Machine learning force fields (MLFFs) are powerful tools for materials modeling, but their performance is often limited by training dataset quality, particularly the lack of rare event configurations. This limitation undermines their accuracy and robustness in long-time and large-scale molecular dynamics simulations. In this work, we present a hybrid MLFF framework that integrates an empirical short-range repulsive potential and demonstrates improved robustness and training efficiency. Using solid electrolyte Li$_7$La$_3$Zr$_2$O$_{12}$ (LLZO) as a model system, we show that purely data-driven MLFFs fail to prevent unphysical atomistic clustering in extended simulations due to inadequate short-range repulsion. In contrast, the hybrid force field eliminates these artifacts, enabling stable long-time simulations, which are critical for studying various properties of LLZO. The hybrid framework also reduces the need for extensive active learning and performs well with just 25 training configurations. By combining physics-driven constraints with data-driven flexibility, this approach is compatible with most existing MLFF architectures and establishes a universal paradigm for developing robust, training-efficient force fields for complex material systems.


\end{abstract}

\maketitle

\section{Introduction}

Machine learning force fields (MLFFs) have demonstrated remarkable capabilities in materials modeling over the past decade~\cite{behler2016perspective,botu2017machine,unke2021machine,ying2025advances}. Empirical force fields, which approximate interatomic interactions through simple predefined functions with built-in physical constraints, risk underfitting in complex materials due to their limited number of empirical parameters. In contrast, MLFFs leverage flexible functional forms and a data-driven approach to capture complex interatomic interactions. A few notable MLFF frameworks include Gaussian approximation potential (GAP)~\cite{bartok2010gaussian}, moment tensor potential~\cite{shapeev2016moment,novikov2020mlip}, deep potential (DP)~\cite{zhang2020dp}, neuroevolution potential (NEP)~\cite{fan2021neuroevolution}, ACE~\cite{drautz2019atomic}, MACE~\cite{Batatia2022mace}, NequIP~\cite{batzner20223}, Allegro~\cite{musaelian2023learning}, CACE~\cite{cheng2024cartesian}, \textit{etc}. These MLFFs are trained on datasets generated from high-accuracy quantum mechanical calculations, such as density functional theory (DFT)~\cite{unke2021machine,ying2025advances}. Once successfully trained, these MLFFs achieve DFT-comparable accuracy, far exceeding empirical force fields while maintaining computational costs orders of magnitude lower than first-principles methods. The combination of high accuracy and low computational cost makes MLFFs well-suited for atomistic modeling in complex material systems, particularly in molecular dynamics (MD) simulations over extended spatial and temporal scales~\cite{unke2021machine}.

As in a data-driven approach, constructing a training dataset that efficiently captures representative atomic configurations is a central challenge in the training of MLFFs. Current strategies primarily involve two approaches: manually perturbing near-equilibrium structures and performing short-time MD simulations integrated with active learning frameworks. However, despite achieving high accuracy on the training set, MLFFs may still fail to generalize beyond the training set sporadically. Consequently, during long-time and/or large-scale MD simulations, close-distance high-energy rare events absent from the training data may arise, exceeding the extrapolation capability of MLFFs, potentially causing simulation breakdown or unphysical results~\cite{finkbeiner2024generating}. The most common remedy for these issues is to extend the training process, such as continuing the iterative active learning loops~\cite{novikov2020mlip,zhang2020dp}. However, this comes at the cost of longer training times and higher computational costs, while the robustness of the retrained force field in long-time MD simulations is still not guaranteed.

As a central component of all-solid-state batteries, ceramic-based solid electrolytes are a class of inorganic materials that typically exhibit fast ionic diffusion and disordered sublattices~\cite{bachman2016inorganic}. Understanding, optimizing, and designing these solid electrolyte materials is crucial for advancing next-generation solid-state batteries~\cite{manthiram2017lithium,zhao2020designing,janek2023challenges}. Since high ionic conductivity is the most characteristic feature of these materials, force fields capable of accurately characterizing fast ion diffusion behavior are an essential prerequisite for reliable atomistic modeling in these material systems~\cite{wang2023frustration,geng2024elucidating,yan2024impact}. These requirements impose strict accuracy criteria on force fields. First, force fields must be capable of accurately describing ion migrations in solids, which are `rare events' in MD simulations. Second, disordered mobile-ion sublattices necessitate force fields capable of generalizing across the diverse atomic configurations sampled during MD simulations. 

In this work, we demonstrate that a hybrid framework integrating physical constraints from empirical potentials with the flexible fitting capability of data-driven MLFFs exhibits significant advantages in both model robustness and training efficiency. Specifically, by incorporating empirical short-range potentials, such as Ziegler-Biersack-Littmark (ZBL) potential~\cite{ziegler1985stopping}, enhances MLFFs robustness by preventing unphysical atomic clustering and catastrophic breakdowns in long-time MD simulations. Besides, this hybrid approach significantly improves training efficiency and lowers the training cost. Using a solid electrolyte material Li$_7$La$_3$Zr$_2$O$_{12}$ (LLZO) as our model system, we comprehensively demonstrate why MLFFs obtained through the conventional data-driven training process fail to prevent unphysical atomic clustering in long-time MD simulations. In contrast, MLFF integrated with ZBL potential~\cite{liu2023large} effectively mitigates this issue, in which the empirical ZBL potential provides a physically driven short-range interaction barrier that prevents unphysical atomic clustering. We show that when ZBL potential is incorporated, even for LLZO, a quaternary system with complex phase transitions and long-range ionic diffusion behaviors, a training set with as few as 25 configurations suffices to achieve reliable performance. Moreover, the integration of ZBL potential reduces the iterative loop of active learning from 13 iterations to 3, significantly improving training efficiency. These performances are difficult or impossible to achieve using the standard MLFF frameworks with conventional training approaches. 

By thoroughly studying the performance and training of MLFFs within the LLZO system, we unveil the critical role of incorporating physics constraints on short-range interactions using empirical potentials. Our approach provides a simple yet broadly applicable strategy to enhance MLFF robustness and training efficiency and can potentially be applied across diverse material systems and reduce computational costs. Our work paves the way for the development of more efficient and robust MLFF training, providing guidance for applying MLFFs to the study of complex material systems.

\raggedbottom
\section{Methods}

\noindent{\textbf{MLFF frameworks}} We employ the neuroevolution potential (NEP)  generation four ~\cite{song2024general} as our baseline framework to construct MLFF for LLZO. NEP is a highly efficient MLFF framework with sufficient simulation accuracy, as demonstrated in our prior studies in the LLZO system~\cite{yan2024impact}. 

Short range Li$^+$-Li$^+$ interactions play a critical role in the concerted ion migration and phase transition behaviors in LLZO~\cite{he2017origin,bernstein2012origin}
However, configurations containing short interatomic distances are rarely sampled during the short-time MD simulations that are typically used in active learning iterations. As we demonstrate below, the insufficient sampling of atomic configurations with short Li$^+$-Li$^+$ pairs in the training dataset can result in poor extrapolation capability of the trained MLFFs, which eventually leads to few but disastrous events or simulation breakdown in long-time MD simulations.    

To account for short-range repulsive interactions, we adopted the universal ZBL potential to describe the short-range part, as implemented in the hybrid NEP-ZBL model~\cite{liu2023large}. The empirical ZBL potential was originally proposed and has been extensively validated for modeling high-energy atomic collisions between atoms~\cite{ziegler1985stopping}. Hybrid MLFF frameworks combined with ZBL potential have been demonstrated successful to simulate the irradiation damage processes of materials in a few previous works~\cite{wang2019deep,byggmastar2019machine,liu2023large}. For example, Wang \textit{et al.} proposed using the DP-ZBL model to simulate irradiation damage processes in a face-centered cubic aluminum system~\cite{wang2019deep}. Byggmästar \textit{et al.} combined GAP with ZBL potential to study the radiation damage and defects in tungsten~\cite{byggmastar2019machine}. Recently, Liu \textit{et al.} developed a hybrid NEP-ZBL framework and demonstrated its performance in large-scale MD simulation of primary radiation damage in tungsten~\cite{liu2023large}. Since we chose NEP as our baseline model, here we adopted the hybrid NEP-ZBL framework to account for the interatomic interactions in the LLZO system. 

The ZBL potential for a pair of atoms $i$ and $j$ takes the form:
\begin{equation}
    U_{ZBL}(r_{i j}) =\frac{1}{4 \pi \epsilon_0} \frac{Z_i Z_j e^2}{r_{i j}} \phi\left(r_{i j} / a\right)f_c\left(r_{i j}\right),
\end{equation}
where $Z_i$ and $Z_j$ are the atomic numbers of the interacting atoms, $e$ is the elementary charge, and $r_{i j}$ is the interatomic distance. 

The universal screening function $\phi(x)$ is given by:
\begin{equation}
    \begin{split}
        \phi(x) &= 0.18175 e^{-3.19980 x} + 0.50986 e^{-0.94229 x} \\
        &\quad + 0.28022 e^{-0.40290 x} + 0.02817 e^{-0.20162 x},
    \end{split}
\end{equation}
with $x = r_{ij}/a$, where $a$ is the screening length (in units of Angstrom) defined as:
\begin{equation}
    a =\frac{0.46848}{Z_i^{0.23}+Z_j^{0.23}},
\end{equation}

The smooth cutoff function, $f_{\mathrm{c}}\left(r_{i j}\right)$, implemented in NEP-ZBL takes the form:
\begin{equation}
    f_{\mathrm{c}}\left(r_{i j}\right)=\left\{\begin{array}{ll}
1 & r_{i j}<r_{\mathrm{c}}^{\mathrm{a}} \\
\frac{1}{2}\left[1+\cos \left(\pi \frac{r_{i j}-r_{\mathrm{c}}^{\mathrm{a}}}{r_{\mathrm{c}}^{\mathrm{b}}-r_{\mathrm{c}}^{\mathrm{a}}}\right)\right] & r_{\mathrm{c}}^{\mathrm{a}} \leq r_{i j} \leq r_{\mathrm{c}}^{\mathrm{b}}, \\
0 & r_{i j}>r_{\mathrm{c}}^{\mathrm{b}}
\end{array}\right.
\end{equation}
where $r_{\mathrm{c}}^{\mathrm{a}}$ and $r_{\mathrm{c}}^{\mathrm{b}}$ are the inner and outer cutoff radii of ZBL potential, respectively. Herein, we use inner and outer cutoff radii of 0.9 Å and 1.8 Å, respectively. Within the inner cutoff, the interactions are fully described by the ZBL potential. Between the inner and outer cutoffs, the interactions gradually transit from the ZBL potential to the NEP description. Outside the outer cutoff, the interactions are fully described by the NEP potential.

Note that since our goal is to prevent catastrophic MD simulation breakdowns using an empirical short-range repulsive potential, the empirical parameters of the ZBL potential, or even the functional form itself, are not particularly critical.

\vspace{0.5cm}

\noindent{\textbf{Machine learning force fields training.}}
We utilized a validated MLFF for LLZO from our previous work~\cite{yan2024impact} as the standard reference, termed NEP$_{std}$. A control MLFF model NEP$_{802}$ was redeveloped via an active learning strategy and converged within 13 iterations, resulting in a training set of 802 structures. To optimize computational efficiency, the energies, forces, and stresses for the NEP$_{802}$ dataset were predicted by the validated NEP$_{std}$ model using \textit{calorine} package~\cite{lindgren2024calorine}, thereby avoiding costly DFT calculations while maintaining accuracy. Based on this dataset, we also trained another model NEP$_{802}$-ZBL, incorporating the ZBL potential to improve short-range interactions. Moreover, we tried to reduce this training set by applying the farthest point sampling method implemented in \href{https://github.com/bigd4/PyNEP}{PyNEP} package, and selected a representative subset of 25 structures, which was used to train the NEP$_{25}$-ZBL model. The entire active learning process was completed through the workflow implemented in \href{https://github.com/zhyan0603/GPUMDkit}{GPUMDkit}.

\vspace{0.5cm}

\noindent{\textbf{Molecular dynamics simulations.}} MD simulations were performed to investigate the phase transition, radial distribution function (RDF), and mean squared displacement (MSD) in LLZO by using \textit{Graphics Processing Units Molecular Dynamics} (GPUMD)~\cite{fan2017efficient}. All simulations were performed using $4\times4\times4$ supercell containing 12288 atoms under the isothermal-isobaric (NPT) ensemble with Martyna-Tuckerman-Tobias-Klein integrators and a timestep of 1 fs. To study the tetragonal-to-cubic phase transition, the system was first equilibrated for 100 ps, followed by cycling the temperature between 600 K and 1200 K over a total simulation time of 2 ns. During the heating process, the system started from 600 K and was heated up to 1200 K at a rate of 300 K/ns. The cooling process went through three stages, with a slower cooling rate near the critical temperature to obtain more accurate results (1200--950 K at a rate of 500 K/ns; 950--850 K at a rate of 100 K/ns; 850--600 K at a rate of 500 K/ns). For the RDF calculations, a 100 ps NPT simulation was performed at each temperature, as the system was gradually heated up. MSDs were calculated at 800 K and 1000 K for 1 ns under NPT ensemble.

\section{Results and Discussion}

\begin{figure*}[]
\centering
\includegraphics[width=0.98\linewidth]{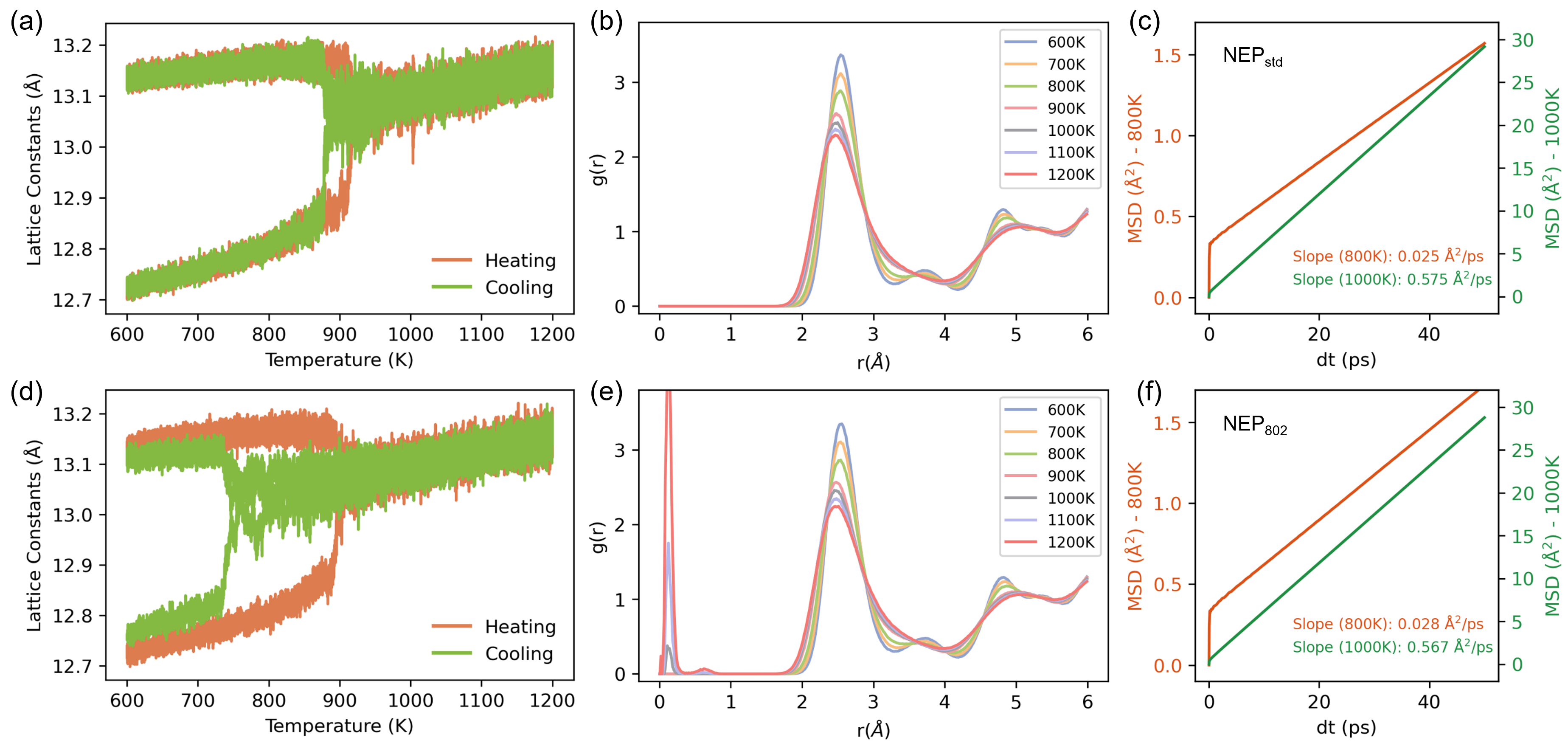}
\caption{(a-c) show the various properties calculated using NEP$_{std}$, including (a) the lattice evolution of LLZO during heating and cooling processes, (b) the radial distribution function of Li$^+$-Li$^+$ at 600 – 1200 K, and (c) the mean square displacements of Li$^+$ ions over correlation times at 800 K and 1000 K, while (d-f) show the corresponding results calculated using NEP$_{802}$.}
\label{fig:fig1}
\end{figure*}

We first established a standard force field NEP$_{std}$ for LLZO as our baseline model. This model was trained on 2024 structures (containing 1978 stoichiometric cubic and tetragonal structures and 46 non-stoichiometric structures with Li-O Schottky defects). This NEP$_{std}$ model, thoroughly validated in our previous work~\cite{yan2024impact} for accurately describing various properties of LLZO including lattice parameters, phase transition behaviors, thermal expansion coefficients, and ion transport properties, served as the surrogate model for DFT and the reference point for our performance benchmark. Herein, partial training hyperparameters were adjusted due to version dependencies and practical considerations to achieve higher training and simulation efficiency (Tab.~S1). The root-mean-square errors (RMSE) of energies, forces, and stresses for the training set are 0.74 meV/atom, 80.31 meV/Å, and 0.089 GPa for NEP$_{std}$ (Fig.~S1).

Next, we try to redevelop an LLZO force field using conventional data-driven active learning strategies. The seed training set includes 100 structures, which were generated through random perturbation of the original tetragonal LLZO structure using the dpdata package~\cite{zeng2023deepmd} with a 5\% cell variation, 0.2 Å atomic displacement, and a `uniform' perturbation style. Afterwards, an active learning strategy was employed to expand the training set and obtain more representative structures. Four samples based on different initial structures were employed to perform MD simulations at different temperatures (100--1200K) and durations (20--500ps) (see Tab.~S2 for details). In this process, we performed descriptor-based farthest point sampling to obtain representative structures, and the final training set contains 802 structures. To optimize computational efficiency, the energies, forces, and stresses for these 802 structures were predicted by the validated NEP$_{std}$ model, thereby avoiding costly DFT calculations while maintaining accuracy. The RMSE of energies, forces, and stresses for the training set are 0.22 meV/atom, 24.96 meV/Å, and 0.037 GPa for NEP$_{802}$ (Fig.~S2). The reduced RMSE in NEP$_{802}$ likely results from the reduced noise in NEP$_{std}$ compared to the original DFT data.

To evaluate the performance of our NEP models, we conducted a comprehensive comparison between NEP$_{std}$ and NEP$_{802}$. \textbf{Figure~\ref{fig:fig1}} presents their comparative performance in describing phase transitions, atomic local environments, and ionic diffusion properties of LLZO. The NEP$_{std}$ model successfully captured the tetragonal-to-cubic phase transition behavior, as shown in Fig.~\ref{fig:fig1}(a). During the heating process, the phase transition occurred around 900 K, where the tetragonal structure transformed into a cubic structure, consistent with previous literature reports~\cite{wang2015phase,chen2015study,yan2024impact}. During the cooling process, the transition temperature shifted to approximately 880 K due to a small thermal hysteresis.

RDF calculations, presented in Fig.~\ref{fig:fig1}(b), reveal the temperature-dependent evolution of local atomic structures. As temperature increases, we observed a gradual broadening and height reduction of the first RDF peak, indicating increasing thermal vibration and an order-disorder transition in the Li sublattice. These RDF trends align well with previous reports~\cite{jalem2015effects}. The MSD calculations, shown in Fig.~\ref{fig:fig1}(c), reflect the Li$^+$ diffusion properties. Using a $4\times4\times4$ supercell containing 3584 Li$^+$ ions over a 1 ns simulation period, we obtained statistically robust time-averaged results. The MSD curves rapidly reached convergence, yielding slopes of 0.025 \AA$^2$/ps and 0.575 \AA$^2$/ps at 800 K and 1000 K, respectively.

In contrast, while NEP$_{802}$ accurately describes the phase transition during the heating process, it showed notable deficiencies in other aspects. As illustrated in Fig.~\ref{fig:fig1}(d), during the cooling process, the cubic phase transforms into the tetragonal phase at about 740 K, and the lattice of the tetragonal phase is discernibly different from that in the heating process. The origin of this discrepancy becomes apparent in the RDF analysis (Fig.~\ref{fig:fig1}(e)), where abnormal peaks appear at distances less than 1 Å (about 0.13 Å and 0.65 Å) in high-temperature simulations (above 900 K). These peak indicate unphysical Li$^+$-Li$^+$ clustering, a phenomenon absent in the NEP$_{std}$ simulations. Meanwhile, the MSD calculations (Fig.~\ref{fig:fig1}(f)) show slight difference compared to NEP$_{std}$ results. These findings demonstrate that while NEP$_{802}$ partially captures the behaviors of LLZO, its tendency to produce unphysical Li$^+$-Li$^+$ clustering at high temperatures leads to discrepancies in phase transition behavior. Such unphysical clustering, although statistically rare, can still occur in long-time MD simulations, leading to unphysical predictions on various properties.

When encountering unexpected predictions or failure in long-time and/or large-scale MLMD simulations, the most conventional remedy is to continue the iterative active learning strategy and expand the training set. However, as we reveal in the LLZO system, catastrophic failures can stem from the deficiency of short-range configurations, which are typically rare events in MD simulations. Unfortunately, sampling rare events in MD simulations is extremely difficult and takes a lot of effort. In a large system like the $4\times4\times4$ supercell of LLZO, 1 ns simulation captures states similar to a single unit cell over 64 ns. Running such a long-time simulation requires significant training resources and time, which is impractical to achieve in the training process. Also, with a 10 fs output interval, a 64 ns trajectory produces about 6.4 million frames, making the descriptor-based sampling method very difficult. Saving frames less, however, might neglect rare events entirely. Therefore, running very long simulations in active learning just to catch a few rare events would be expensive and impractical.

Our observation reveals that the under-fitted short-range repulsion interaction in the trained MLFF is due to the lack of close-distance high-energy structures, which are intrinsically difficult to be captured by conventional MD sampling strategies in the active learning process. In fact, the RDF analysis reveals that, apart from the unphysical Li$^+$-Li$^+$ clustering at distances below 1 Å, NEP$_{802}$ produces results remarkably similar to the NEP$_{std}$ at larger distances. This observation suggests that NEP$_{802}$ could potentially achieve performance comparable to NEP$_{std}$ through a simple addition of appropriate short-range repulsive interactions. 

\begin{figure}[t]
\centering
\includegraphics[width=0.98\linewidth]{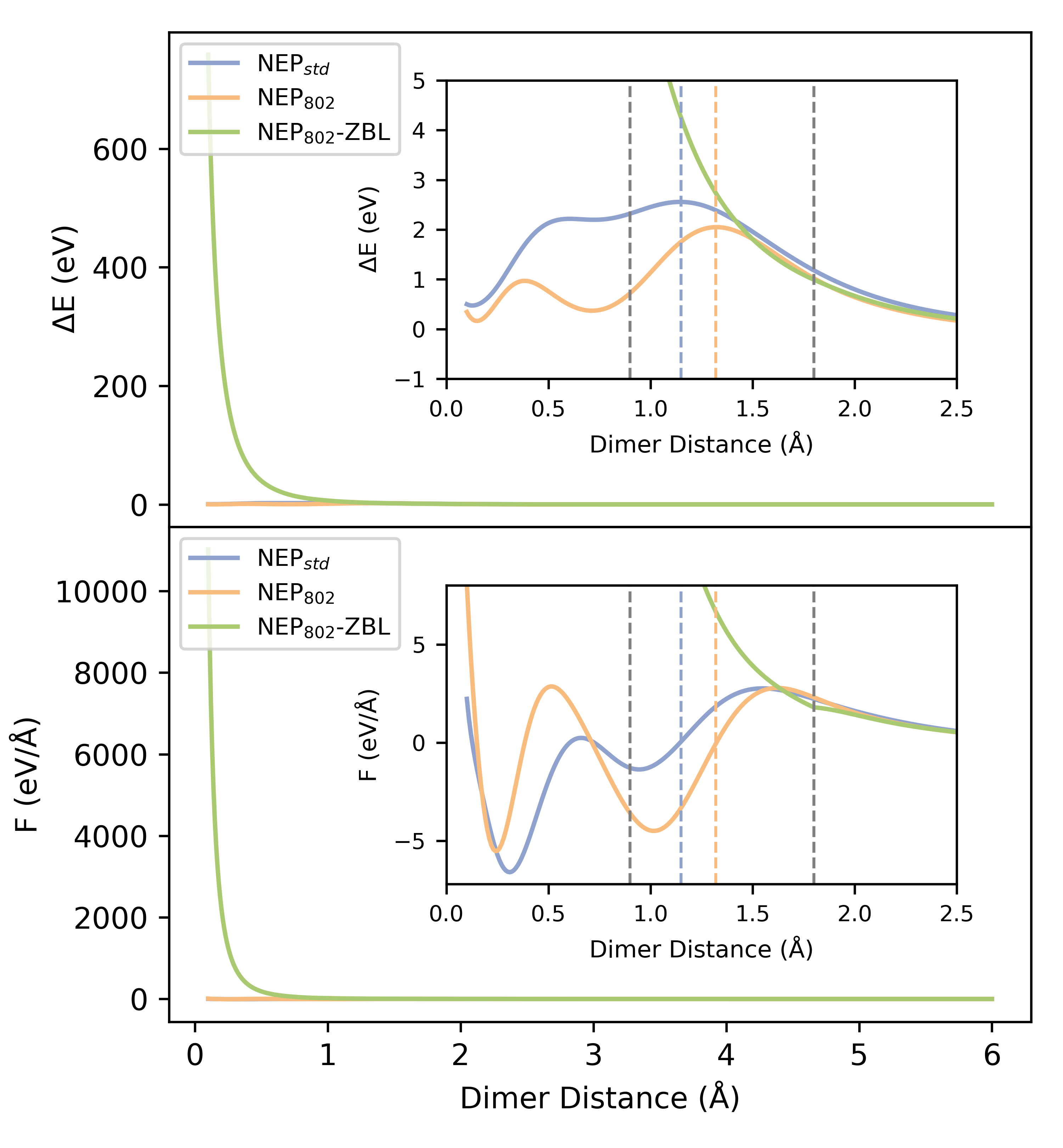}
\caption{The potential energy (upper panel) and force (lower panel) as functions of Li-Li dimer distance, with the inset showing a zoom-in of the short-range region. The gray dashed lines indicate the cutoff positions of ZBL-inner = 0.9 Å and ZBL-outer = 1.8 Å. The blue and yellow dashed lines correspond to the breakdown distances. The energy and force values were shifted to zero at a dimer distance of 6 Å for better comparison.}
\label{fig:fig2}
\end{figure}

To overcome the deficiency, a remedial approach is to use an empirical potential to handle the short-range interactions below a certain cutoff distance, while leaving MLFF to learn the interactions outside the cutoff distance. Herein, we use a general ZBL potential to describe the short-range repulsive interaction. This hybrid NEP-ZBL model has been implemented in the GPUMD and NEP program~\cite{fan2022gpumd} and has been successfully applied to simulate the large-scale radiation damage in tungsten~\cite{liu2023large}. In fact, we find that this hybrid framework can improve the robustness of MLFFs, even though it was originally proposed for the simulation of radiation damage~\cite{wang2019deep,byggmastar2019machine,liu2023large}.

We retrained a NEP-ZBL model using the dataset of NEP$_{802}$, called NEP$_{802}$-ZBL, and compared the performance of NEP$_{std}$, NEP$_{802}$, and NEP$_{802}$-ZBL. To evaluate the short-range interactions, we placed two Li atoms forming a dimer in a cubic box with a box length of 30 Å. \textbf{Figure~\ref{fig:fig2}} illustrates the potential energy (upper panel) and force (lower panel) as functions of Li-Li dimer distance, with the inset focusing on the short-range region. For comparative purposes, both the energy and force values were shifted to zero at a distance of 6 Å. It is important to note that none of these force fields were explicitly trained on isolated Li-Li dimer configurations; rather, they learned Li$^+$-Li$^+$ interactions within the LLZO framework. Therefore, while these results may not be directly comparable to DFT calculations of isolated dimers, they provide valuable insights into the capability of each model to capture short-range interactions.

As the Li-Li distance decreases, the energy curves show different behavior. While all three NEP models show an increase in initial energy as the dimer distance decreases, NEP$_{802}$ exhibits a significant energy drop when the dimer distance is below 1.32 Å. We call this the breakdown distance, that is, the dimer distance when Li-Li undergoes unphysical short-range attractive interaction. At a closer dimer distance, it exhibits two local minima of potential energy, located at about 0.15 Å and 0.7 Å, which explains why the abnormal Li$^+$-Li$^+$ aggregation peaks at about 0.13 Å and 0.65 Å appeared in the previous RDF analysis (Fig.~\ref{fig:fig1}(e)). In contrast, NEP$_{std}$ maintains the repulsive trend until a shorter distance of 1.15 Å, before showing a similar decreasing trend. This explains why the NEP$_{std}$ model does not exhibit Li$^+$-Li$^+$ aggregation in the simulations. However, NEP$_{std}$ also exhibits unphysical short-range Li-Li attractive interactions, indicating that at higher temperatures or longer simulation times, NEP$_{std}$ may also experience unphysical atomic aggregation similar to NEP$_{802}$. Both NEP$_{std}$ and NEP$_{802}$ models give unreliable results in the short-range region, where the potential energy surface becomes arbitrary. In contrast, NEP$_{802}$-ZBL, augmented with the empirical repulsive potential, consistently maintains the expected repulsive behavior throughout the short-range region. The force curves further highlight these differences. NEP$_{802}$ exhibits a decrease in repulsive force when the dimer distance is below 1.62 Å, while NEP$_{std}$ shows a decrease in repulsive force at a dimer distance below 1.55 Å. The NEP-ZBL model, on the other hand, exhibits complete repulsion in the short-range region.

\begin{figure}[]
	\centering
	\includegraphics[width=0.98\linewidth]{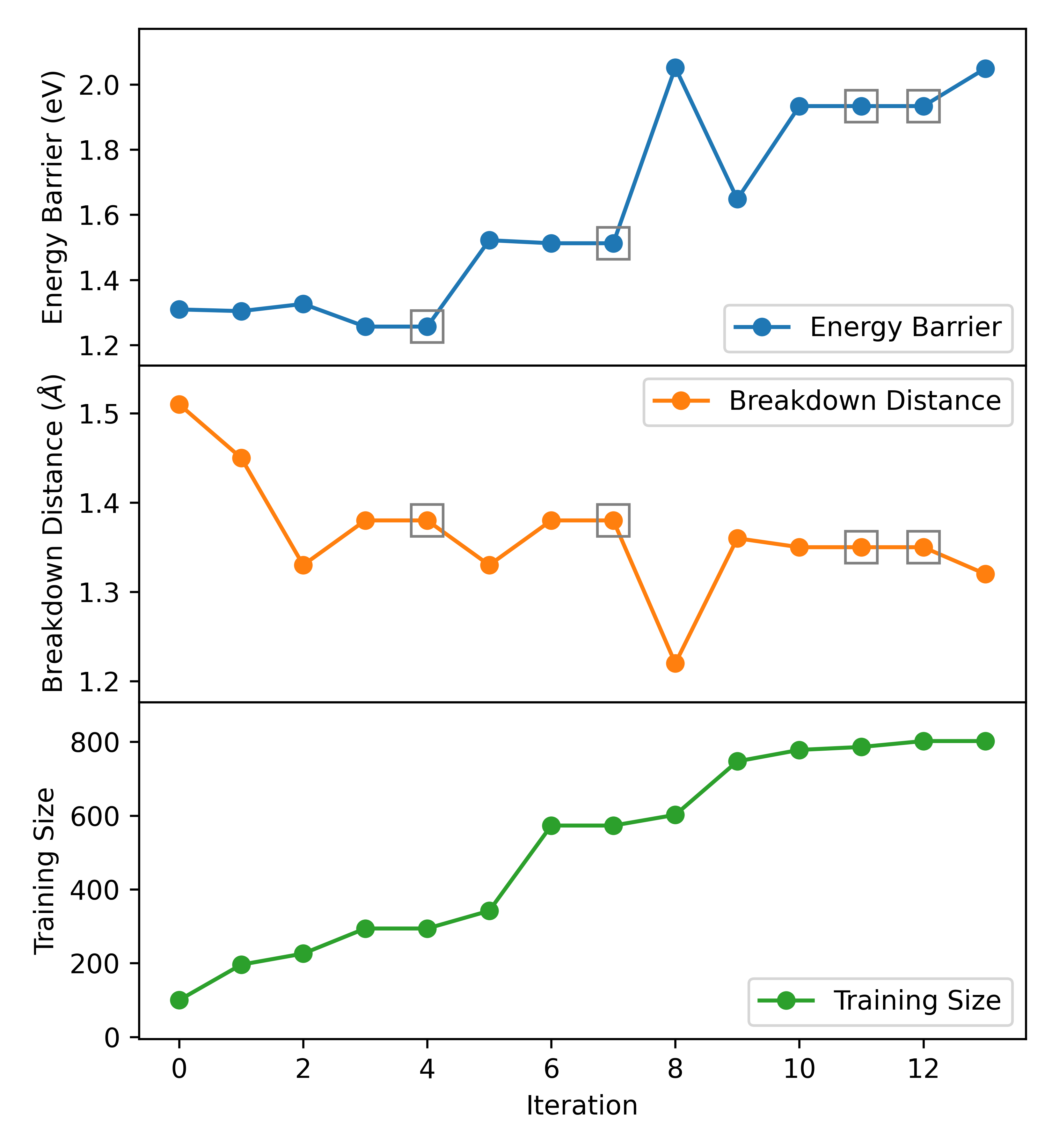}
	\caption{Evolution of energy barriers, breakdown distances, and training set size over active learning iterations. Grey hollow squares indicate that the force field was not updated in this iteration and used data from the previous round.}
	\label{fig:fig3}
\end{figure}

The arbitrary behavior of NEP$_{802}$ in the short-range region introduces a very interesting question: how do barriers and breakdown distances evolve during active learning? \textbf{Figure}~\ref{fig:fig3} shows that with the progress of active learning iteration, the energy barriers show an increasing trend, while the breakdown distances show a decreasing trend. These trends suggest that the iterative MD sampling processes encounter some structures with close atomic distances, thereby gradually enhancing the learning of short-range interactions. However, this learning process is extremely inefficient. After 13 rounds of active learning, the energy barrier only rises from 1.31 eV to 2.05 eV, and the breakdown distance decreases from 1.51 Å to 1.32 Å. This inefficiency stems from the rarity of such structures in the MD sampling process; achieving higher energy barriers and lower breakdown distances may require longer simulations to capture a sufficient number of short-range interaction configurations. It is worth noting that our NEP$_{std}$ training set contains some high-pressure structures achieved by compressing the lattice, which may explain its better ability to learn repulsive interactions. Even so, it still exhibits a breakdown distance of 1.15 Å, only 0.17 Å shorter than that of NEP$_{802}$. This highlights the significant challenges of learning short-range interactions through active learning iterations.

\begin{figure*}
	\centering
	\includegraphics[width=0.98\linewidth]{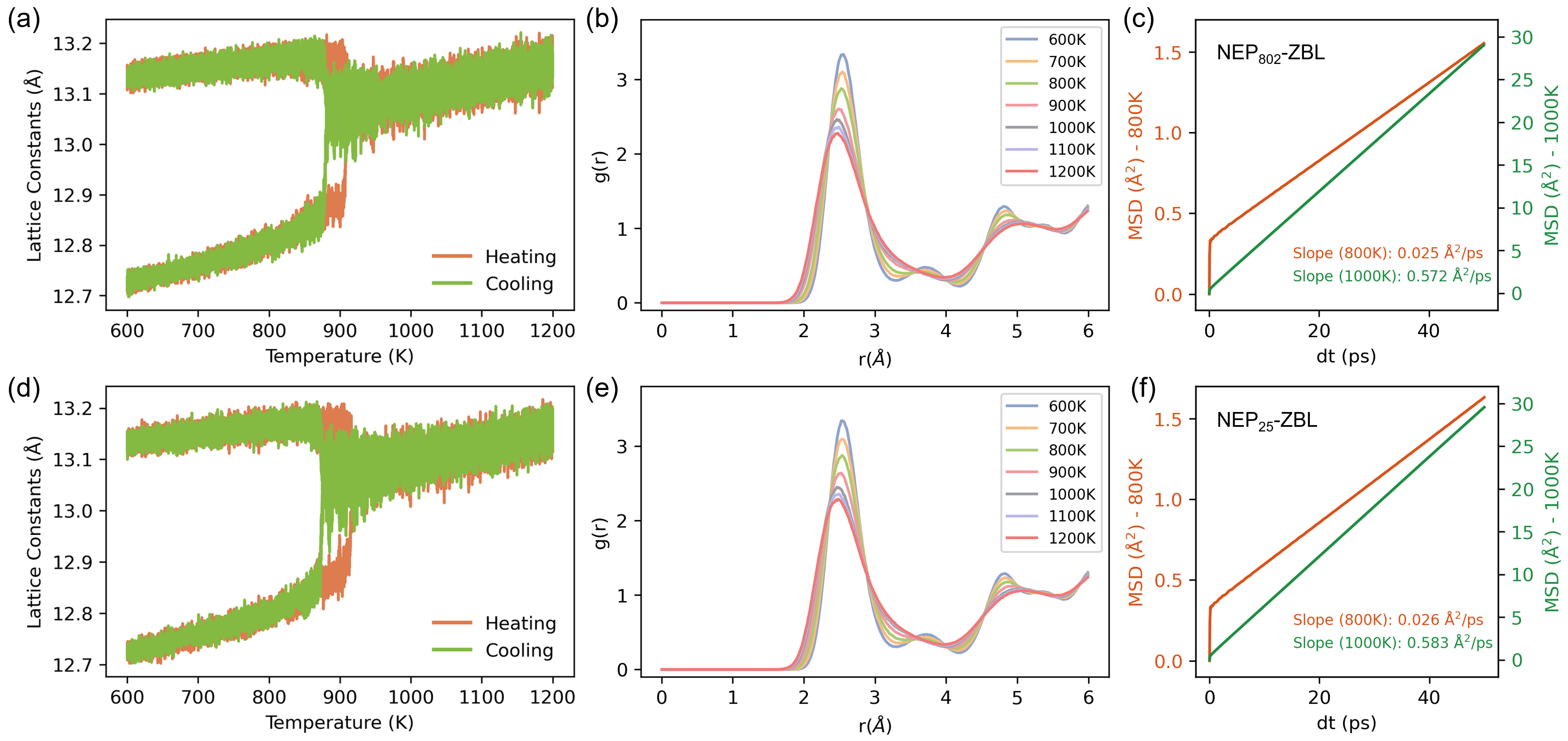}
	\caption{(a-c) show the various properties calculated using NEP$_{802}$-ZBL, including (a) the lattice evolution of LLZO during heating and cooling processes, (b) the radial distribution function of Li$^+$-Li$^+$ at 600 – 1200 K, and (c) the mean square displacements of Li$^+$ ions over correlation times at 800 K and 1000 K, while (d-f) show the corresponding results calculated using NEP$_{25}$-ZBL.}
	\label{fig:fig4}
\end{figure*}

To prevent the arbitrary behavior of MLFFs in the short-range region, we can integrate the empirical repulsive potential into the MLFFs to take over the short-range interactions. NEP-ZBL resolves the limitations of describing short-range interactions by introducing an insurmountable barrier, thus avoiding unphysical atomic aggregation. \textbf{Figure~\ref{fig:fig4}}(a-c) presents the performance of NEP$_{802}$-ZBL, evaluated using the same metrics as previously discussed: phase transitions, RDF, and MSD. The results demonstrate significant improvements after incorporating the ZBL repulsive potential. The NEP$_{802}$-ZBL model accurately captures the tetragonal-to-cubic phase transition, with the critical temperature occurring around 900 K during the heating process and around 880 K during the cooling process, matching the behavior of the NEP$_{std}$ model. The RDF analyses show that the NEP$_{802}$-ZBL model successfully maintains proper short-range repulsion, effectively eliminating the unphysical Li$^+$-Li$^+$ clustering observed in the NEP$_{802}$ model. The MSD calculations reveal that NEP$_{802}$-ZBL achieves nearly identical diffusion behavior to the NEP$_{std}$ model, with slopes of 0.025 Å$^2$/ps and 0.572 Å$^2$/ps at 800 K and 1000 K, respectively. 

Based on our experience with the LLZO system and in handling short-range interactions, we further explored the possibility of reducing the training set size while maintaining the performance of the force field. We refined the training set of NEP$_{802}$ into a smaller one with only 25 configurations and trained another NEP$_{25}$-ZBL model (See SI for the training errors of NEP$_{25}$-ZBL model). As shown in Fig.~\ref{fig:fig4}(d-f), despite using a tiny training set, the NEP$_{25}$-ZBL model successfully reproduces various properties of LLZO, including the tetragonal-to-cubic phase transition during both heating and cooling processes, as well as RDF and MSD results. While the MSD from NEP$_{25}$-ZBL shows slight deviations, with slopes of 0.026 Å$^2$/ps and 0.583 Å$^2$/ps at the corresponding temperatures, these values remain remarkably close to those from the NEP$_{std}$ model. The success of the NEP$_{25}$-ZBL model, trained on only 25 structures, suggests that the local atomic environments in the quaternary LLZO system may be less complex than previously assumed. 

The remarkable performance of NEP$_{25}$-ZBL also raises another intriguing question: how can a force field trained on such a tiny dataset with just 25 configurations accurately reproduce the results of the NEP$_{std}$ model? To address this question, we performed Uniform Manifold Approximation and Projection (UMAP) analysis on the Li$^+$ descriptors from three datasets: the NEP$_{std}$ dataset containing 1978 stoichiometric LLZO configurations, NEP$_{802}$ dataset with 802 configurations, and NEP$_{25}$ dataset with 25 configurations. In the NEP framework, these descriptors are high-dimensional vectors comprising radial and angular components that characterize the local atomic environments. The construction details of these descriptors can be found in reference~\cite{fan2022gpumd}.

\begin{figure}[b]
	\centering
	\includegraphics[width=0.90\linewidth]{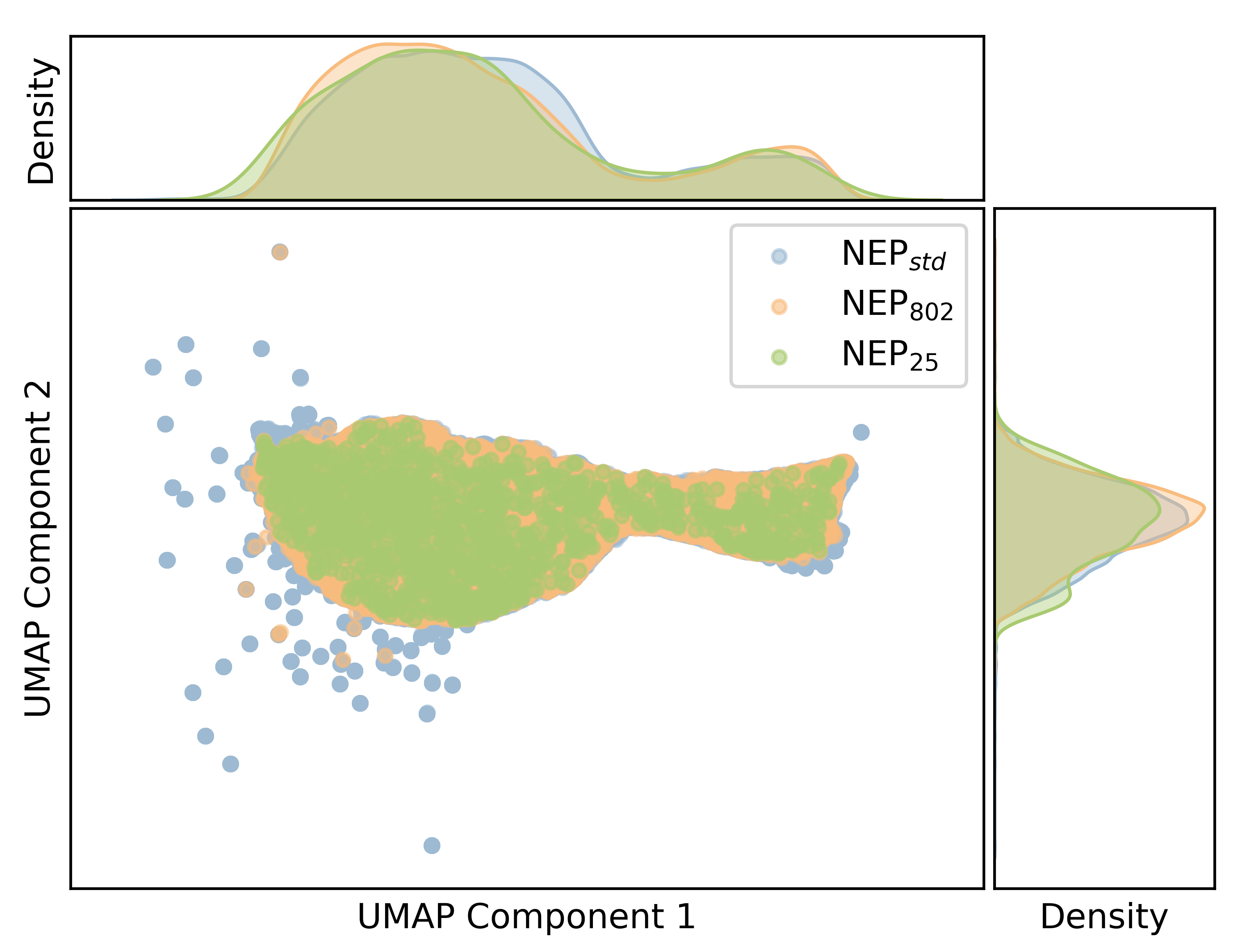}
	\caption{Uniform Manifold Approximation and Projection of Li$^+$ high-dimensional descriptors in different training sets.}
	\label{fig:fig5}
\end{figure}

Based on our hyperparameter choice, we generated a 30-dimensional description vector for each atom. Using the nonlinear unsupervised UMAP method, a widely adopted technique for analyzing high-dimensional data, we discovered that the datasets of NEP$_{802}$ and NEP$_{25}$ cover nearly the entire descriptor space sampled in the datasets of NEP$_{std}$ (\textbf{Figure}~\ref{fig:fig5}). This substantial coverage indicates that most Li$^+$ local environments are successfully captured even with significantly fewer training configurations. The small regions of uncovered Li$^+$ descriptor space (blue scatters) likely correspond to high-pressure configurations present in the NEP$_{std}$ dataset but absent in both NEP$_{802}$ and NEP$_{25}$ datasets. The linear algorithm, Principal Component Analysis, yields results similar to those of UMAP (Fig.~S7). 


The performance of NEP$_{25}$-ZBL (Fig.~\ref{fig:fig4}) and the analysis of UMAP (Fig.~\ref{fig:fig5}) demonstrate that once the short-range interactions are properly handled, the atomic interactions in LLZO can be easily captured in a limited training set. This observation implies that by incorporating an empirical short-range repulsive potential, it not only improves the robustness of the trained MLFF, but can also significantly improve the training efficiency and reduce the  training set size. 

We performed a comparative analysis on the training efficiency of MLFFs, with and without empirical short-range potential. Table~\ref{tab:iterations} lists the performance of NEP and NEP-ZBL models during exploration iterations of active learning. Both models underwent three loops of active learning, with MD sampling performed for 500 ps at temperature ranges of 100–400 K, 500–800 K, and 800–1200 K, respectively. Structures were selected from the MD trajectories using the farthest point sampling method to augment the training dataset. 

\begin{table}[!t]
    \centering
    \caption{Exploration iterations during the training of LLZO force fields with NEP and NEP-ZBL frameworks. The Failure Ratio (FR) represents the fraction of unphysical structures within the trajectory relative to the total structure counts. Unphysical structures are filtered out during the sampling process.}
    \label{tab:iterations}
        \begin{tabular}{@{}ccccccc@{}}
        \toprule
        \multirow{2}{*}{Iter.}& \multirow{2}{*}{t (ps)} & \multirow{2}{*}{Temperature} & \multicolumn{2}{c}{NEP-ZBL} & \multicolumn{2}{c}{NEP} \\ \cmidrule(l){4-7} 
                              &                         &                              & N$_{samp}$     & FR (\%)     & N$_{samp}$       & FR (\%)      \\ \cmidrule(r){1-7}
        1                     & 500                     & 100-400K                     & 12             & 0\%         & 10               & 0\%          \\
        2                     & 500                     & 500-800K                     & 51             & 0\%         & 22               & 19.9\%       \\
        3                     & 500                     & 800-1200K                    & 44             & 0\%         & 29               & 57.9\%       \\ \bottomrule
        \end{tabular}
\end{table}

For the model trained with the ZBL potential, a total of 207 structures were obtained after three iterations, resulting in the NEP$_{207}$-ZBL model. As expected, this model successfully reproduced the phase transition, RDF, and MSD of the LLZO system (Fig.~S8). In contrast, the model trained without the ZBL potential exhibited significant instability. In the second iteration, it showed a failure rate of 19.9\%, meaning that 19.9\% of the structures in the simulated trajectories exhibited unphysical atomic clustering. In the third iteration, this failure rate even rose to 57.9\%. These results indicate that the default data-driven MLFF framework (without ZBL repulsive potential) cannot sustain such an aggressive training strategy involving rapid temperature increase and extended simulation times, as it will generate a substantial number of unphysical trajectory structures.

The hybrid framework of MLFF and empirical repulsive potential proves to be a low-cost yet powerful tool for preventing simulation catastrophes, particularly when training datasets are not yet comprehensive. It enables us to use more aggressive sampling strategies, reduce active learning iteration loops, accelerate force field development, and help avoid unnecessary expansion of the training sets. By providing a robust repulsive wall, the empirical short-range repulsive potential effectively addresses the challenge of short-range interactions. Although events involving strong repulsive forces are relatively rare during simulations, the absence of adequate energy barriers can lead to significant simulation catastrophes. These results highlight the critical importance of proper short-range repulsion in MD simulations.

\section{Conclusion}

In this work, we introduce a hybrid framework that integrates the empirical ZBL potential into MLFFs, significantly enhancing their robustness and training efficiency for complex material systems like the quaternary LLZO solid electrolyte. Our systematic analysis reveals that purely data-driven MLFFs, such as the NEP$_{802}$ model, often fail to capture short-range repulsive interactions due to the rarity of close-distance configurations in training datasets, leading to erroneous extrapolations of short-range interactions and unphysical atomic clustering during long-time MD simulations. By incorporating the ZBL potential, which is grounded in the physical principles of atomic repulsion, our hybrid NEP-ZBL models effectively eliminate these unphysical configurations, ensuring robust simulations in predicting phase transitions, local atomic environments, and ionic transport properties of LLZO. Remarkably, the NEP$_{25}$-ZBL model, trained on only 25 configurations, achieves performance comparable to the NEP$_{std}$ model trained on nearly 2000 configurations, demonstrating that strategic inclusion of physical constraints can drastically reduce training data requirements. UMAP analysis further confirms that even a small dataset can cover the essential Li$^+$ descriptor space, which explains the decently well performance of NEP$_{25}$-ZBL model. The physical constraints introduced by the ZBL repulsive potential compensate for the extrapolation limitations of data-driven approaches, particularly in handling data-sparse parts, such as rare events or extreme configurations. Moreover, this hybrid approach reduces active learning iterations from 13 to 3, greatly improving the training efficiency of MLFFs. 

Our work provides valuable insights for the development of robust and efficient machine learning potentials for complex materials. It suggests that careful consideration of fundamental physical constraints can significantly improve the robustness of MLFFs and reduce the data requirements without compromising accuracy. By unifying physics-driven constraints with data-driven flexibility, our framework offers a generalizable paradigm for developing robust, training-efficient MLFFs. The methodology demonstrated here, while focused on LLZO as a case study, offers a promising strategy for developing robust and efficient force fields for other solid-state electrolytes and complex materials. Future work could explore the applicability of this approach to incorporate other physical constraints, such as long-range dispersion corrections, that might also enhance the robustness and training efficiency of MLFFs in different material systems.

\vspace{0.5cm}

\section*{Data availability}
Supporting information and partial source files are available on the GitHub repository at \href{https://github.com/zhyan0603/SourceFiles}{\textcolor{blue}{https://github.com/zhyan0603/SourceFiles}}. All other source files, including MD sampling trajectories and various output files of GPUMD and NEP, can be obtained from the corresponding author upon request.

\section*{Code availability}
The source code and tutorials for our GPUMDkit are available at \href{https://github.com/zhyan0603/GPUMDkit}{\textcolor{blue}{https://github.com/zhyan0603/GPUMDkit}}.

\section*{Acknowledgements}
We acknowledge the support from the Research Center for Industries of the Future (RCIF) and the High-Performance Computing Center (HPC) at Westlake University.

\section*{Author contributions}
Y. Zhu and Z. Yan conceived and designed the research; Y. Zhu and Z. Fan guided the research; Z. Yan performed the simulations; Z. Yan, Z. Fan, and Y. Zhu wrote the paper.

\section*{Competing Interests}
The authors declare no competing interests.

\bibliography{references}
\end{document}